\documentclass[%
 preprint,
 nofootinbib,
 amsmath,amssymb,
 aps,
 a4paper
]{revtex4-1}

\usepackage{graphicx}
\usepackage{dcolumn}
\usepackage{bm}
\usepackage{slashed}
\usepackage{color}
\usepackage{comment}
\usepackage{tikz-feynhand}
\usepackage{physics}

\begin{document}

\title{Search for WIMPs at future $\mu^+\mu^+$ colliders}

\author{Hajime Fukuda}
\email{hfukuda@hep-th.phys.s.u-tokyo.ac.jp}
\affiliation{Department of Physics, The University of Tokyo, Tokyo 113-0033, Japan}

\author{Takeo Moroi}
\email{moroi@hep-th.phys.s.u-tokyo.ac.jp}
\affiliation{Department of Physics, The University of Tokyo, Tokyo 113-0033, Japan}

\author{Atsuya Niki}
\email{niki@hep-th.phys.s.u-tokyo.ac.jp}
\affiliation{Department of Physics, The University of Tokyo, Tokyo 113-0033, Japan}

\author{Shang-Fu Wei}
\email{sfwei@hep-th.phys.s.u-tokyo.ac.jp}
\affiliation{Department of Physics, The University of Tokyo, Tokyo 113-0033, Japan}


\begin{abstract}
Weakly interacting massive particles (WIMPs) with electroweak charges, such as the wino and the Higgsino, stand out as natural candidates for dark matter in the universe. 
In this paper, we study the search for WIMPs at future multi-TeV $\mu^+\mu^+$ colliders. We investigate both the direct production search of WIMPs through the mono-muon channel and the indirect search through quantum corrections in elastic $\mu^+\mu^+$ M{\o}ller scattering. We find that the indirect search has an advantage over the direct search with sufficient luminosities, $\mathcal{O}(\text{ab}^{-1})$, and low systematic uncertainties, $\lesssim 0.3\,\%$. This advantage arises due to the weaker mass dependence observed in the indirect search in comparison to direct production methods. The advantage is further enhanced if the initial muon beams are polarized.
Specifically, we demonstrate that the indirect search method can detect the thermal mass target for the wino and the Higgsino for $\sqrt{s} = 6\,\text{TeV}$ and $2\,\text{TeV}$ (with $\sqrt{s}$ being the center of mass energy), respectively, with $10\,\text{ab}^{-1}$, an $80\,\%$ polarized beam and an accuracy of $0.1\,\%$. 
Our findings illuminate the potential of future high-energy $\mu^+\mu^+$ colliders in advancing our understanding of dark matter.
\end{abstract}

\maketitle


\section{Introduction}
The presence of dark matter in the universe stands as one of the most compelling pieces of evidence for physics beyond the standard model\,\cite{Planck:2018vyg}.
Despite this certainty, the true nature of dark matter remains enigmatic, leading to the exploration of various proposed candidates\,\cite{Feng:2010gw}. Among these, the weakly interacting massive particle (WIMP)\,\cite{Lee:1977ua} is a prominent candidate and has been extensively studied due to its ability to yield the correct dark matter abundance through the freeze-out mechanism.

The supersymmetric extension of the standard model (SSM) provides an intriguing framework for WIMP dark matter\,\cite{Jungman:1995df} since the lightest supersymmetric particle is stable with conserved $R$-parity\,\cite{Farrar:1978xj}. In particular, the wino or the Higgsino, the supersymmetric partner of the SU$(2)_{L}$ gauge boson or the Higgs boson, respectively, contains electrically neutral components and are the natural candidates for dark matter in the SSM. If other supersymmetric particles are much heavier, the electroweak gauge interaction determines the relic abundance; the correct relic abundance is thermally produced for the $\sim 2.7\,\text{TeV}$ wino or the $\sim 1\,\text{TeV}$ Higgsino\,\cite{Hisano:2006nn,Cirelli:2007xd}. Such a scenario is known as the pure wino or Higgsino, respectively. These scenarios are one of the most motivated targets for WIMP searches, referred to as ``the thermal mass targets''.
The pure wino and Higgsino, a $\text{SU}(2)_L$ triplet fermion and a $\text{SU}(2)_L$ doublet fermion with the hypercharge $1/2$, respectively, can be interpreted as the simplest candidates of the dark matter; the neutral component of such electroweak multiplets can, without any other interaction than the electroweak interaction, automatically be stable and be the WIMP dark matter. It is thus studied from a bottom-up point of view, called the minimal dark matter\,\cite{Cirelli:2005uq,Cirelli:2007xd}.

Collider experiments present a promising avenue for WIMP search.
Much experimental effort has been made in the ATLAS\,\cite{ATLAS:2013ikw,ATLAS:2017oal,ATLAS:2022rme} and CMS\,\cite{CMS:2014gxa,CMS:2018rea,CMS:2020atg} experiment at the LHC. The WIMP is also one of the main targets of future colliders\,\cite{Arkani-Hamed:2015vfh,EuropeanStrategyforParticlePhysicsPreparatoryGroup:2019qin,Han:2020uak}.
Among these, muon colliders emerge as particularly promising candidates due to their capability to achieve high-energy collisions with substantial luminosities. Muons, being heavier than electrons, experience suppressed synchrotron radiation, enabling their acceleration in circular accelerators and enhancing both luminosity and energy. Additionally, muons, being elementary particles unlike protons, carry the entire energy for collision.
This unique feature positions high-energy $\mu^+\mu^-$ colliders as potent tools for probing WIMPs with good significance\,\cite{Han:2020uak}.

A significant challenge faced in the development of muon colliders lies in creating bright $\mu^\pm$ sources with low emittance\,\cite{Delahaye:2019omf,MAP_Delahaye:2013jla,MICE:2019jkl,LEMMA_Antonelli:2015nla}. 
For $\mu^+$ beams, however, there is an existing technique\,\cite{Kondo:2018rzx} to achieve sufficiently low emittance\,\cite{MAP_Delahaye:2013jla}; it uses the ionization of muonium atoms and can be applied to only positive muons. This fact opens the possibility of considering high-energy $\mu^+\mu^+$ colliders as a more realistic option than $\mu^+\mu^-$ colliders\,\cite{Heusch:1995yw,Hamada:2022mua,Endo:2022imj,Hamada:2022uyn,Lichtenstein:2023iut,Fridell:2023gjx,Dev:2023nha}. Ref.\,\cite{Hamada:2022mua}, for example, pointed out that a multi-TeV $\mu^+\mu^+$ collider is possible with currently available technologies. It is thus of interest to examine the capability of future $\mu^+\mu^+$ colliders to discover the WIMP.

At the LHC\,\cite{ATLAS:2013ikw,ATLAS:2017oal,ATLAS:2022rme,CMS:2014gxa,CMS:2018rea,CMS:2020atg} or future hadron\,\cite{Arkani-Hamed:2015vfh} or $e^+e^-$ colliders\,\cite{EuropeanStrategyforParticlePhysicsPreparatoryGroup:2019qin}, the most promising technique for the WIMP searches is to find the disappearing track of the charged component of the WIMP.
However, at muon colliders, detecting disappearing tracks may be challenging because muon colliders notoriously suffer from large beam-induced backgrounds\,\cite{MuonCollider:2022ded}. 
Muons' finite lifetime leads to decay in flight, resulting in numerous decays per meter. Decay products hitting materials produce additional particles reaching detectors.
With the timing cuts, the hit densities in the trackers are reduced but still expected to be $\mathcal{O}(100)/\text{cm}^2$ at the innermost layer\,\cite{MuonCollider:2022ded}, which is one-order larger than the HL-LHC\,\cite{ATLAS:2017svb}.
We thus need to develop detailed designs of the detector and improve the track reconstruction algorithms\,\cite{Capdevilla:2021fmj,Han:2020uak}.


In this paper, we shift our focus to both (i) the direct production of WIMPs involving an additional particle and (ii) indirect measurement through radiative corrections induced by WIMPs on the elastic M{\o}ller scattering cross section of muons at future $\mu^+\mu^+$ colliders. The pure Higgsino, the wino, and the $\text{SU}(2)$ 5-plet Majorana fermion serve as our primary targets for this study. 
We will find that the indirect search method is more efficient than the direct search method for probing heavier WIMP with sufficient luminosity because the mass dependence of the indirect methods is weaker than the on-shell production. 
We will show that multi-TeV $\mu^+\mu^+$ colliders can reveal the WIMP with masses $\mathcal{O}(\text{TeV})$. In particular, our findings reveal that, with integrated luminosity of $10\,\text{ab}^{-1}$, $80\,\%$ polarization, and $0.1\,\%$ accuracy, $\mu^+\mu^+$ colliders can probe the thermal mass target of the $1$ TeV Higgsino and $2.7$ TeV wino for the center-of-mass energy of $2\,\text{TeV}$ and $6\,\text{TeV}$, respectively.

The paper is organized as follows. In Sec.\,\ref{sec:indirect}, we review the indirect effect of the WIMP on the elastic scattering between $\mu^+$s. We then discuss the direct production search of the WIMP at $\mu^+\mu^+$ colliders in Sec.\,\ref{sec:direct}. In Sec.\,\ref{sec:result}, we present our analysis methods and show our results. Finally,  the conclusion and discussion are in Sec.\,\ref{sec:summary}.

\section{WIMP effect on muon elastic scattering}
\label{sec:indirect}
\setlength{\feynhandlinesize}{1pt}
\begin{figure}[t]
    \centering
    \begin{tikzpicture}
        \begin{feynhand}
            \vertex (i1) at (-2.5,1.5) {$\mu^+$};
            \vertex (i2) at (-2.5,-1.5) {$\mu^+$};
            \vertex (f1) at (2.5,1.5) {$\mu^+$};
            \vertex (f2) at (2.5,-1.5) {$\mu^+$};
            \vertex (m1) at (0,1.5);
            \vertex (m2) at (0,-1.5);
            \vertex (m11) at (0,0.7);
            \vertex (m12) at (0,-0.7);
            \propag [fer] (m1) to (i1);
            \propag [fer] (m2) to (i2);
            \propag [fer] (f1) to (m1);
            \propag [fer] (f2) to (m2);

            \propag [pho] (m1) to [edge label = $\gamma/Z$](m11);
            \propag [bos] (m12) to [edge label =$\gamma/Z$](m2);
            \propag [fer] (m11) to [half right, looseness=1.7](m12);
            \propag [fer] (m12) to [edge label = $\Tilde{\chi}$, half right, looseness=1.7](m11);
        \end{feynhand}
    \end{tikzpicture}
    \caption{The WIMP effect on the $\mu^+\mu^+$ elastic scattering. $\Tilde{\chi}$ is the WIMP, which interacts with standard model particles only via gauge bosons.}
    \label{fig:diagram}
\end{figure}
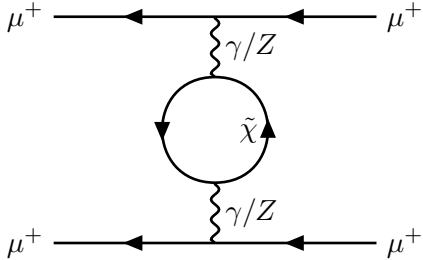

In this section, we summarize the indirect effect of the WIMPs\,\cite{Harigaya:2015yaa} on the $\mu^+\mu^+$ elastic scattering. Let us consider a fermionic WIMP charged under the standard model $\text{SU}(2)_L\times \text{U}(1)_Y$. The new particle affects the $\mu^+\mu^+$ elastic cross section through the vacuum polarization of the standard model gauge bosons. The Feynman diagram for the $t$-channel gauge boson exchange process is shown in Fig.~\ref{fig:diagram}, where the WIMP effect appears in the photon and $Z$ boson vacuum polarizations. By integrating the WIMP out, the effective Lagrangian is obtained as
\begin{align}
    \mathcal{L} = \mathcal{L}_{\textrm{SM}} + \frac{C_{WW}}{4}W_{\mu\nu}^a\Pi(-D^2/m^2)W^{a\mu\nu} + \frac{C_{BB}}{4}B_{\mu\nu}\Pi(-\partial^2/m^2)B^{\mu\nu} + \cdots,
    \label{eq:eft lagrangian}
\end{align}
where $\mathcal{L_{\textrm{SM}}}$ is the standard model Lagrangian, $W_{\mu\nu}^a$ ($B_{\mu\nu}$) is the field strength tensor for $\textrm{SU}(2)_{L}$ ($\textrm{U}(1)_{Y}$) gauge group, $D$ is the covariant derivative and $m$ is the mass of the WIMP. The coefficients $C_{WW}$ and $C_{BB}$ for the $\textrm{SU}(2)_{L}\ n$-plet WIMP with hypercharge $Y$ are given by
\begin{align}
    C_{WW} &= \kappa\cdot\frac{2}{3}g^2(n^3-n),\\
    C_{BB} &= \kappa\cdot8(g \tan\theta_W)^2 nY^2,
\end{align}
where $g$ is the gauge coupling for $\textrm{SU}(2)_{L}$, $\theta_W$ is the Weinberg angle, and $\kappa$ is $1\ \qty(\frac{1}{2})$ for Dirac (Majorana) fermion. $\Pi(x)$ denotes the loop function of the vacuum polarization due to the WIMP loop,
\begin{align}
    \Pi(x) = \frac{1}{16\pi^2}\int_{0}^{1}\dd{y} y(1-y)\log\qty(\frac{m^2-xy(1-y)m^2 }{\mu^2}),
\end{align}
where $\mu$ is the renormalization scale. We set the scale as $\mu=m$ so that $\Pi(x\ll 1)=0$.

In order to evaluate the effect of the WIMP on the $\mu^+\mu^+$ elastic scattering, let us expand the scattering amplitude perturbatively with the coupling constant.
The standard model process contributes to the tree-level amplitude. The tree-level scattering amplitude for $\mu^+(p_1)\mu^+(p_2)\rightarrow\mu^+(p_3)\mu^+(p_4)$ is obtained as
\begin{align}
    \mathcal{M}_{\textrm{tree}} &= \qty[\Bar{v}(p_3)\gamma^\mu P_{h_1}v(p_1)] \qty[\Bar{v}(p_4)\gamma_\mu P_{h_2}v(p_2)] \sum_{V=\gamma,Z}\frac{C^{V}_{h_1}C^{V}_{h_2}}{t-m_V^2} - (u\textrm{-channel}),
\end{align}
where $h_i=L,R$ is the helicity of the initial $i$-th muon, $P_h$ is the projection operator onto the helicity $h$ eigenstate, $v(p)$ is the $\mu^+$ wave function for momentum $p$, and
$C^V_h$ is 
given by 
$C^\gamma_{L} = C^\gamma_{R} = -e$, $C^Z_{L} = \frac{g}{c_W}s_W^2$, and $C^Z_{R} = \frac{g}{c_W}(-1/2+s_W^2)$.

The effect of the WIMP appears in the 1-loop diagram. By using Eq.~\eqref{eq:eft lagrangian}, the 1-loop WIMP contribution is obtained as
\begin{align}
    \mathcal{M}_{\textrm{WIMP}} = \qty[\Bar{v}(p_3)\gamma^\mu P_{h_1}v(p_1)] \qty[\Bar{v}(p_4)\gamma_\mu P_{h_2}v(p_2)] \sum_{V,V'=\gamma,Z}C^{V}_{h_1}C^{V'}_{h_2}&\frac{C_{VV'}t\Pi(t/m^2)}{(t-m_V^2)(t-m_{V'}^{2})} \nonumber\\
    &- (u\textrm{-channel}),
\end{align}
where $C_{\gamma\gamma}=s_W^2C_{WW}+c_w^2C_{BB},\ C_{ZZ}=c_W^2C_{WW}+s_w^2C_{BB}$ and $C_{\gamma Z} = C_{Z\gamma}=s_W c_W (C_{WW}-C_{BB})$. The $t$ and $u$ dependence of $\Pi$ induces the shape-difference of angular distribution between the differential cross section with and without the WIMP.

The differential cross section can be also expanded pertubatively.
The effect of the WIMP appears in the differential cross section as the interference between the tree-level and 1-loop amplitude.
The ratio of the tree-level differential cross sections, $\dd{\sigma}_{\textrm{tree}}/\dd{\cos\theta}$, to the WIMP contribution to the differential cross section, $\dd{\sigma}_{\textrm{WIMP}}/\dd{\cos\theta}$, is given by
\begin{align}
\label{eq:deltatheta}
    \Delta(\theta) \equiv \frac{\dd{\sigma}_{\textrm{WIMP}}/\dd{\cos\theta}}{\dd{\sigma}_{\textrm{tree}}/\dd{\cos\theta}} = \frac{2\Re (\mathcal{M}_{\textrm{tree}}\mathcal{M}_{\textrm{WIMP}}^{*})}{|\mathcal{M}_{\textrm{tree}}|^2},
\end{align}
where we define $\theta$ as the scattering angle from the beam axis. 


The ratio $\Delta(\theta)$ depends on the polarization of the initial muons,
because the $\text{U}(1)_Y\times\text{SU}(2)_L$ charges of right-handed muons and left-handed muons are different.
For simplicity, we assume that the polarizations of the muon beams from both sides are the same; let $P_{\mu^+}$ be the polarization of the initial muons with $P_{\mu^+}=+1$ and $-1$ corresponding to right- and left-handed, respectively. The differential cross section is written as
\begin{align}
    \frac{d\sigma}{d\cos\theta} = \frac{1}{4}&\left[(1+P_{\mu^+})(1+P_{\mu^+})\frac{d\sigma_{RR}}{d\cos\theta}+(1-P_{\mu^+})(1-P_{\mu^+})\frac{d\sigma_{LL}}{d\cos\theta}+\right. ,\nonumber\\
    &\left.(1+P_{\mu^+})(1-P_{\mu^+})\frac{d\sigma_{RL}}{d\cos\theta}+(1-P_{\mu^+})(1+P_{\mu^+})\frac{d\sigma_{LR}}{d\cos\theta}\right],
\end{align}
where $\frac{d\sigma_{IJ}}{d\cos\theta}\ (I,J=L,R)$ is the differential cross section between $I$-handed and $J$-handed muons.
As the $\text{SU}(2)_L$ coupling is larger for WIMPs and only the right-handed $\mu^+$s enjoy the $\text{SU}(2)_L$ charge, we expect that the larger $P_{\mu^+}$ is, the larger $\Delta(\theta)$ is.

\begin{figure}[t]
\begin{center}
\includegraphics[scale=0.7]{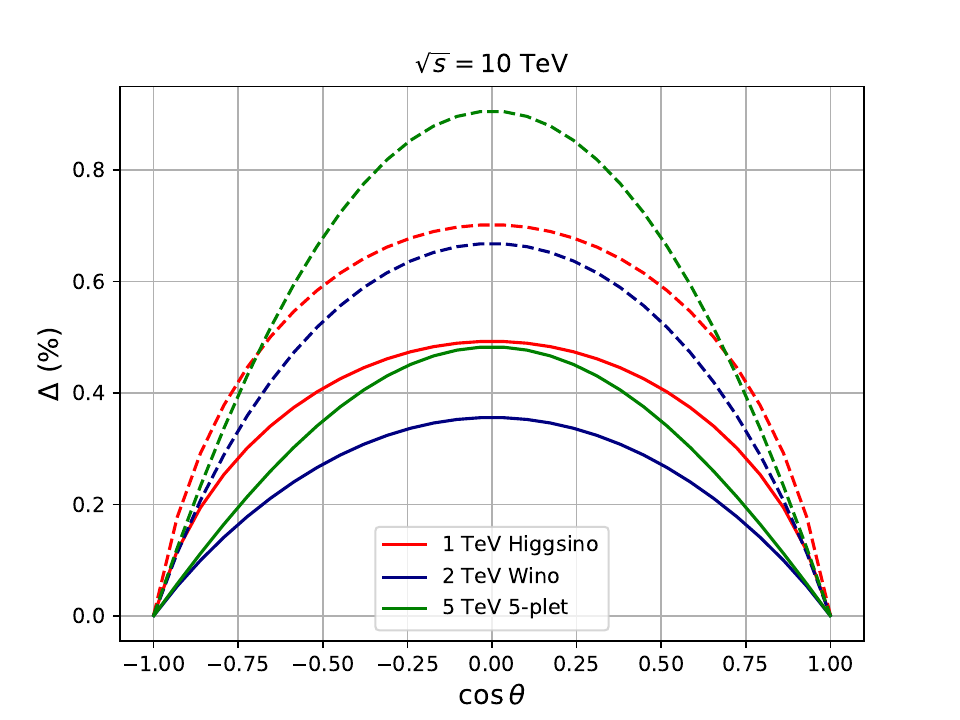}
\end{center}
\caption{The angular distribution of the cross section ratio $\Delta$ in the case of Higgsino, Wino, and 5-plet minimal dark matter. The center of mass energy is $\sqrt{s}=10$ TeV and the $\mu^+$ polarization is either $P_{\mu^+}=0$ (solid line) or $0.8$ (dashed line).}
\label{fig:cross_ratio}
\end{figure}

Fig.~\ref{fig:cross_ratio} shows the ratio $\Delta(\theta)$ in several cases: $1$ TeV Higgsino ($n=2,\ Y=\frac{1}{2}$, Dirac fermion), $2$ TeV Wino ($n=3,\ Y=0$, Majorana fermion) and $5$ TeV 5-plet minimal dark matter ($n=5,\ Y=0$, Majorana fermion) with $\sqrt{s}=10$ TeV (where $\sqrt{s}$ is the center of mass energy of the $\mu^+\mu^+$ system). The muon polarization is assumed to be either $P_{\mu^+}=0$ (solid line) or $0.8$ (dashed line).
As expected, when the initial $\mu^+$s are right-handed, $\Delta$ is enhanced. For example, for the wino or $5$-plet, which do not have the hypercharge, the muon polarization enhances the WIMP contribution by the factor $(1+P_{\mu^+})(1+P_{\mu^+})$ compared to the one with the unpolarized beam. On the other hand, the background cross section is approximately proportional to $\frac{1}{2} (1+P_{\mu^+})(1+P_{\mu^+})$ because numerically $\frac{d\sigma_{RR}}{d\cos\theta} \simeq \frac{d\sigma_{LL}}{d\cos\theta} >  \frac{d\sigma_{LR}}{d\cos\theta}, \frac{d\sigma_{RL}}{d\cos\theta}$ in the standard model. The ratio $\Delta$ is, therefore, expected to be enhanced by factor $\sim2$ when the muons are polarized as $P_{\mu^+}\sim 1$ as observed in Fig.~\ref{fig:cross_ratio}.

From the figure, we can see that the WIMP affects the shape of the angular distribution of the elastic scattering. The ratio $\Delta$ has the largest value in the most transverse region, \textit{i.e.} $\cos\theta=0$ because the vacuum polarization is large with large momentum transfer. On the other hand, the differential cross section is smaller for the larger angle scattering and the statistics are better in the forward regions.
This behavior implies that it is important for the indirect search of the WIMP at $\mu^+\mu^+$ colliders to analyze the angular distribution of $\mu^+\mu^+$ elastic scattering. We further discuss the detail in Sec\,\ref{sec:result}. 

\section{Direct production}
\label{sec:direct}


We next examine the direct production search for the WIMP at $\mu^+\mu^+$ colliders. For the search at $\mu^+\mu^-$ colliders\,\cite{Han:2020uak}, two types of signals, mono-$\mu$ and mono-$\gamma$, are the main channels. We examine these two channels at $\mu^+\mu^+$ colliders. Due to the lack of the Drell-Yan (DY) process for $\mu^+\mu^+$ collisions, unlike $\mu^+\mu^-$ collisions, we will see that the mono-$\mu$ channel is the most sensitive channel for the detection of the WIMP dark matter.

We use {\sc Madgraph5}\,\cite{Alwall:2014hca} event generator to perform Monte-Carlo simulations and compare the number of signals with the standard model backgrounds. To create the model file for {\sc Madgraph5}, we use {\sc FeynRules 2.0}\,\cite{Alloul:2013bka}. We have checked the consistency of the model file with Ref.\,\cite{Han:2020uak} and our analytic calculations.

For the mono-$\mu$ channel, the signal processes are followings;
\begin{align}
    \mu^+ \mu^+ &\to \mu^+ \bar{\nu}_\mu \chi \bar{\chi} \\
    \gamma \mu^+ &\to \mu^+ \chi \bar{\chi}
\end{align}
Here, we adopt the improved Weizs\"acker-Williams approximation\,\cite{Frixione:1993yw} to include photons as the initial partons. The representative Feynman diagrams of the signals are almost the same as ones at $\mu^+\mu^-$ colliders, as shown in Fig.\,\ref{fig:signal_monomu}.

\begin{figure}[t]
\begin{minipage}{0.3\textwidth}
\begin{center}
\begin{tikzpicture}
\begin{feynhand}
            \vertex (i1) at (-2.0,2.0) {$\gamma$};
            \vertex (i2) at (-2.0,0.0) {$\mu^+$};
            \vertex (n1) at (2.0,2.0) {$\nu$};
            \vertex (n2) at (2.0,1.0) {$\bar{\nu}$};
            \vertex (f2) at (2.15, 0.0) {$\mu^+$};

            \vertex (m11) at (-0.5,0.0);
            \vertex (m12) at (0.4,0.0);
            \vertex (m13) at (0.8, 1.5);
            
            \propag [pho] (i1) to (m11);
            \propag [pho] (m12) to [edge label =$Z$](m13);
            \propag [fer] (n2) to (m13);
            \propag [fer] (m13) to (n1);
            \propag [fer] (m11) to (i2);
            \propag [fer] (m12) to (m11);
            \propag [fer] (f2) to (m12);
        \end{feynhand}
\end{tikzpicture}
\end{center}
\end{minipage}
\hspace{0.1\textwidth}
\begin{minipage}{0.3\textwidth}
\begin{center}
\begin{tikzpicture}
\begin{feynhand}
            \vertex (i1) at (-2.0,2.0) {$\gamma$};
            \vertex (i2) at (-2.0,0.0) {$\mu^+$};
            \vertex (n1) at (2.0,1.0) {$\nu_\mu$};
            \vertex (n2) at (2.15,2.0) {$\mu^+$};
            \vertex (f2) at (2.0, 0.0) {$\bar{\nu}_\mu$};

            \vertex (m11) at (-0.5,0.0);
            \vertex (m12) at (0.4,0.0);
            \vertex (m13) at (0.8, 1.5);
            
            \propag [pho] (i1) to (m11);
            \propag [pho] (m12) to [edge label =$W^+$](m13);
            \propag [fer] (n2) to (m13);
            \propag [fer] (m13) to (n1);
            \propag [fer] (m11) to (i2);
            \propag [fer] (m12) to (m11);
            \propag [fer] (f2) to (m12);
        \end{feynhand}
\end{tikzpicture}
\end{center}
\end{minipage}
    \caption{The representative Feynman diagrams for the mono-$\mu$ backgrounds.}
    \label{fig:bg_monomu}
\end{figure}
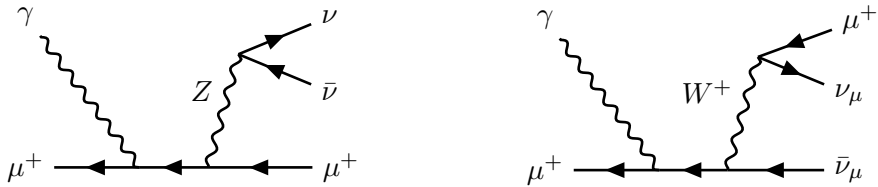

We apply the same cut as Ref.\,\cite{Han:2020uak} for the mono-$\mu$ channel. First, we require one hard muon to be detected; the energy of the muon should be larger than $E_\mu>0.23\sqrt{s}$ with the pseudorapidity $|\eta| < 2.5$. Unlike $\mu^+\mu^-$ colliders, we cannot impose a forward region cut, as the initial particles are identical. We also impose the missing mass cut, $(p_{\mu^+, 1}^\text{in} + p_{\mu^+, 2}^\text{in} - p_{\mu^+}^\text{out})^2 > 4 m_\chi^2$, where $p_{\mu^+, i}^\text{in}$ and $p_{\mu^+}^\text{out}$ are the four-momentum for the initial $i$-th muon and the final hard muon, respectively, and $m_\chi$ is the dark matter mass.
With these cuts, the main standard model background processes are
\begin{align}
    \gamma \mu^+ &\to \mu^+ Z, Z \to \nu \bar{\nu}\\
    \gamma \mu^+ &\to \bar{\nu}_\mu W^+, W^+ \to \nu_\mu \mu^+,
\end{align}
as shown in Fig.\,\ref{fig:bg_monomu}. 

\begin{figure}[t]
\begin{minipage}{0.3\textwidth}
\begin{center}
\begin{tikzpicture}
\begin{feynhand}
            \vertex (i1) at (-2.0,1.1) {$\mu^+$};
            \vertex (i2) at (-2.0,-1.1) {$\mu^+$};
            \vertex (f1) at (2.0,1.5) {$\bar{\nu}_\mu$};
            \vertex (f2) at (2.0,-1.5) {$\mu^+$};
            \vertex (x1) at (1.8, 0.5) {$\bar{\chi}$};
            \vertex (x2) at (1.8, -0.5) {$\chi$};
            \vertex (m1) at (0,1.1);
            \vertex (m2) at (0,-1.1);
            \vertex (m11) at (0,0.5);
            \vertex (m12) at (0,-0.5);
            \propag [fer] (m1) to (i1);
            \propag [fer] (m2) to (i2);
            \propag [fer] (f1) to (m1);
            \propag [fer] (f2) to (m2);

            \propag [pho] (m1) to [edge label'=$W^+$](m11);
            \propag [pho] (m12) to [edge label' =$\gamma/Z$](m2);
            \propag [fer] (m11) to (m12);
            \propag [fer] (x1) to (m11);
            \propag [fer] (m12) to (x2);
        \end{feynhand}
\end{tikzpicture}
\end{center}
\end{minipage}
\hspace{0.1\textwidth}
\begin{minipage}{0.3\textwidth}
\begin{center}
\begin{tikzpicture}
\begin{feynhand}
            \vertex (i1) at (-2.0,1.5) {$\gamma$};
            \vertex (i2) at (-2.15,-1.5) {$\mu^+$};
            \vertex (f2) at (2.0,-1.5) {$\mu^+$};
            \vertex (x1) at (2.0, 0.5) {$\bar{\chi}$};
            \vertex (x2) at (2.0, -0.5) {$\chi$};
            \vertex (m2) at (0,-1.5);
            \vertex (m11) at (0,0.5);
            \vertex (m12) at (0,-0.5);
            \propag [fer] (m2) to (i2);
            \propag [fer] (f2) to (m2);
            \propag [pho] (i1) to (m11);
            \propag [fer] (m11) to (m12);
            \propag [fer] (x1) to (m11);
            \propag [fer] (m12) to (x2);

            \propag [bos] (m12) to [edge label =$\gamma/Z$](m2);
        \end{feynhand}
\end{tikzpicture}
\end{center}
\end{minipage}
    \caption{The representative Feynman diagrams for the mono-$\mu$ signals. $\chi$ denotes the dark matter.}
    \label{fig:signal_monomu}
\end{figure}
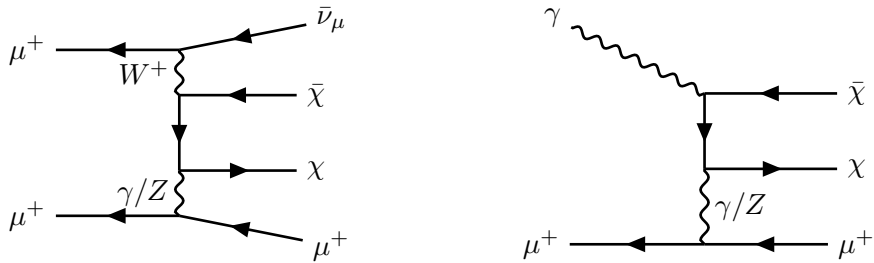

\begin{figure}[t]
\begin{minipage}{0.3\textwidth}
\begin{center}
\begin{tikzpicture}
\begin{feynhand}
            \vertex (i1) at (-2.0,1.5) {$\gamma$};
            \vertex (i2) at (-2.0,-1.5) {$\gamma$};
            \vertex (x1) at (2.0, 1.5) {$\bar{\chi}$};
            \vertex (x2) at (2.0, -1.5) {$\chi$};
            \vertex (po) at (2.0, 0.0) {$\gamma$};

            \vertex (m11) at (0,0.8);
            \vertex (pi) at (0.4, 0.94);
            \vertex (m12) at (0,-0.8);
            \propag [pho] (i1) to (m11);
            \propag [fer] (m11) to (m12);
            \propag [fer] (x1) to (m11);
            \propag [fer] (m12) to (x2);

            \propag [bos] (m12) to (i2);
            \propag [pho] (pi) to (po);
        \end{feynhand}
\end{tikzpicture}
\end{center}
\end{minipage}
\begin{minipage}{0.3\textwidth}
\begin{center}
\begin{tikzpicture}
\begin{feynhand}
            \vertex (i1) at (-2.0,1.5) {$\gamma$};
            \vertex (i2) at (-2.15,-1.5) {$\mu^+$};
            \vertex (f2) at (2.0,-1.5) {$\bar{\nu}_\mu$};
            \vertex (x1) at (2.0, 0.5) {$\chi$};
            \vertex (x2) at (2.0, -0.5) {$\bar{\chi}$};
            \vertex (po) at (2.0, -2.0) {$\gamma$};
            
            \vertex (m2) at (0,-1.5);
            \vertex (m11) at (0,0.5);
            \vertex (m12) at (0,-0.5);
            \vertex (pi) at (-1.0, -1.5);
            
            \propag [fer] (m2) to (i2);
            \propag [fer] (f2) to (m2);
            \propag [pho] (i1) to (m11);
            \propag [fer] (m12) to (m11);
            \propag [fer] (m11) to (x1);
            \propag [fer] (x2) to (m12);

            \propag [bos] (m12) to [edge label' =$W^+$](m2);
            \propag [pho] (pi) to (po);
        \end{feynhand}
\end{tikzpicture}
\end{center}
\end{minipage}
\begin{minipage}{0.3\textwidth}
\begin{center}
\begin{tikzpicture}
\begin{feynhand}
            \vertex (i1) at (-2.0,1.5) {$\mu^+$};
            \vertex (i2) at (-2.0,-1.5) {$\mu^+$};
            \vertex (f1) at (2.0,1.5) {$\bar{\nu}_\mu$};
            \vertex (f2) at (2.0,-1.5) {$\bar{\nu}_\mu$};
            \vertex (x1) at (2.0, 0.7) {$\chi^+$};
            \vertex (x2) at (2.0, -0.7) {$\chi^+$};
            \vertex (po) at (2.0, -2.0) {$\gamma$};
            
            \vertex (m1) at (0,1.5);
            \vertex (m2) at (0,-1.5);
            \vertex (pi) at (-1.0, -1.5);
            \vertex (m11) at (0,0.7);
            \vertex (m12) at (0,-0.7);
            \vertex [crossdot] (mMa) at (0, 0) {};
            \propag [fer] (m1) to (i1);
            \propag [fer] (m2) to (i2);
            \propag [fer] (f1) to (m1);
            \propag [fer] (f2) to (m2);

            \propag [pho] (m1) to [edge label'=$W^+$](m11);
            \propag [pho] (m12) to [edge label' =$W^+$](m2);
            \propag [fer] (mMa) to (m11);
            \propag [fer] (mMa) to (m12);
            \propag [fer] (m11) to (x1);
            \propag [fer] (m12) to (x2);
            \propag [pho] (pi) to (po);
        \end{feynhand}
\end{tikzpicture}
\end{center}
\end{minipage}
    \caption{The representative Feynman diagrams for the mono-$\gamma$ signals. The blob in the rightmost diagram denotes the Majorana mass of the dark matter.}
        \label{fig:signal_monophoton}
\end{figure}
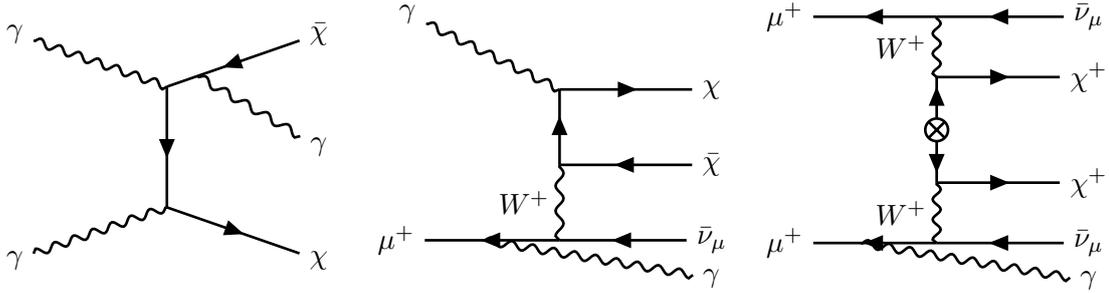

For the mono-$\gamma$ channel, the signal processes are following;
\begin{align}
    \gamma \gamma &\to \chi \bar\chi \gamma \\
    \gamma \mu^+ &\to \chi \bar\chi \bar{\nu}_\mu \gamma \\
    \mu^+ \mu^+ &\to \bar{\nu}_\mu \bar{\nu}_\mu \chi^+ \chi^+.
\end{align}
The representative diagrams are shown in Fig.\,\ref{fig:signal_monophoton}.
We have two main differences in this channel compared with $\mu^+\mu^-$ colliders discussed in Ref.\,\cite{Han:2020uak}.
First, the DY process is unavailable for $\mu^+\mu^-$ colliders. This fact significantly reduces the cross section for the case of heavy WIMP and spoils the advantage of the mono-$\gamma$ channel.
Second, the third mode, $\mu^+ \mu^+ \to \bar{\nu}_\mu \bar{\nu}_\mu \chi^+ \chi^+$, violates the fermion number. This means that the dark matter must be Majorana and the amplitude is proportional to the Majorana mass term. This mode is available for the pure wino or Majorana $5$-plet, but not for the pure Higgsino as the pure Higgsino is Dirac.\footnote{For the Higgsino, a tiny Majorana mass splitting $\sim \mathcal{O}(100)\,\text{keV}$ between the two neutralinos is needed from the phenomenological point of view to avoid the dark matter direct detection constraints\,\cite{Nagata:2014wma}.
}

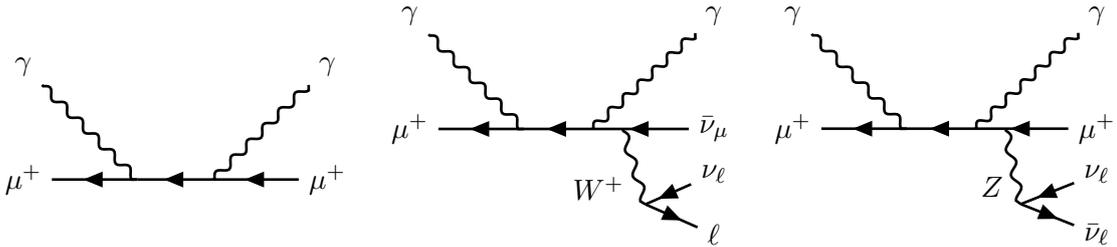
\begin{figure}[t]
\begin{minipage}{0.3\textwidth}
\begin{center}
\begin{tikzpicture}
\begin{feynhand}
            \vertex (i1) at (-2.0,1.5) {$\gamma$};
            \vertex (i2) at (-2.0,0.0) {$\mu^+$};
            \vertex (f1) at (2.0,1.5) {$\gamma$};
            \vertex (f2) at (2.0, 0.0) {$\mu^+$};

            \vertex (m11) at (-0.5,0.0);
            \vertex (m12) at (0.5,0.0);
            \propag [pho] (i1) to (m11);
            \propag [pho] (m12) to (f1);
            \propag [fer] (m11) to (i2);
            \propag [fer] (m12) to (m11);
            \propag [fer] (f2) to (m12);
        \end{feynhand}
\end{tikzpicture}
\end{center}
\end{minipage}
\begin{minipage}{0.3\textwidth}
\begin{center}
\begin{tikzpicture}
\begin{feynhand}
            \vertex (i1) at (-2.0,1.5) {$\gamma$};
            \vertex (i2) at (-2.0,0.0) {$\mu^+$};
            \vertex (f1) at (2.0,1.5) {$\gamma$};
            \vertex (f2) at (2.0, 0.0) {$\bar{\nu}_\mu$};
            \vertex (f3) at (2.0, -0.6) {$\nu_\ell$};
            \vertex (f4) at (2.0, -1.4) {$\ell$};

            \vertex (m11) at (-0.5,0.0);
            \vertex (m12) at (0.4,0.0);
            \vertex (m13) at (0.8,0.0);
            
            \vertex (g1) at (1.1,-1.0);
            
            \propag [pho] (i1) to (m11);
            \propag [pho] (m12) to (f1);
            \propag [fer] (m11) to (i2);
            \propag [fer] (m12) to (m11);
            \propag [fer] (f2) to (m12);
            \propag [fer] (f3) to (g1);
            \propag [fer] (g1) to (f4);

            \propag [pho] (m13) to [edge label'=$W^+$](g1);
        \end{feynhand}
\end{tikzpicture}
\end{center}
\end{minipage}
\begin{minipage}{0.3\textwidth}
\begin{center}
\begin{tikzpicture}
\begin{feynhand}
            \vertex (i1) at (-2.0,1.5) {$\gamma$};
            \vertex (i2) at (-2.0,0.0) {$\mu^+$};
            \vertex (f1) at (2.0,1.5) {$\gamma$};
            \vertex (f2) at (2.0, 0.0) {$\mu^+$};
            \vertex (f3) at (2.0, -0.6) {$\nu_\ell$};
            \vertex (f4) at (2.0, -1.4) {$\bar{\nu}_\ell$};

            \vertex (m11) at (-0.5,0.0);
            \vertex (m12) at (0.4,0.0);
            \vertex (m13) at (0.8,0.0);
            
            \vertex (g1) at (1.0,-1.0);
            
            \propag [pho] (i1) to (m11);
            \propag [pho] (m12) to (f1);
            \propag [fer] (m11) to (i2);
            \propag [fer] (m12) to (m11);
            \propag [fer] (f2) to (m12);
            \propag [fer] (f3) to (g1);
            \propag [fer] (g1) to (f4);

            \propag [pho] (m13) to [edge label'=$Z$](g1);
        \end{feynhand}
\end{tikzpicture}
\end{center}
\end{minipage}
    \caption{The representative Feynman diagrams for the mono-$\gamma$ backgrounds.}
        \label{fig:bg_monophoton}
\end{figure}

For the background of the mono-$\gamma$ channel, as the charge of the initial particles is non-zero, a charged lepton in the final states must be missed. We consider three kinds of process with the charged lepton escaped as the main source of the background as shown in Fig.\,\ref{fig:bg_monophoton}:
\begin{align}
    \mu^+\gamma &\to \mu^+\gamma \\
    \mu^+\gamma &\to \mu^+\gamma Z, Z\to\nu\bar{\nu} \\
    \mu^+\gamma &\to \bar{\nu}\gamma W^+, W^+ \to \ell^+\nu.
\end{align}
We assume a charged lepton cannot be reconstructed if it is in the forward region, $|\eta| > 2.5$, or the transverse momentum $p_T$ is smaller than $10\,\text{GeV}$, which is taken from the default cuts of {\sc Delphes3}\,\cite{deFavereau:2013fsa} for the ATLAS and CMS detector simulation. We require these conditions for the charged leptons in these processes.
To reduce the background, we require $|\eta| < 1.5$ and $p_T > 10\,\text{GeV}$ for $\gamma$. We also require the missing mass cut as the mono-$\mu$ channel, $(p_{\mu^+, 1}^\text{in} + p_{\mu^+, 2}^\text{in} - p_{\gamma}^\text{out})^2 > 4 m_\chi^2$, where $p^\gamma$ is the four-momentum for the photon.

\begin{figure}[t]
\begin{center}
    \centering
        \includegraphics[scale=0.6]{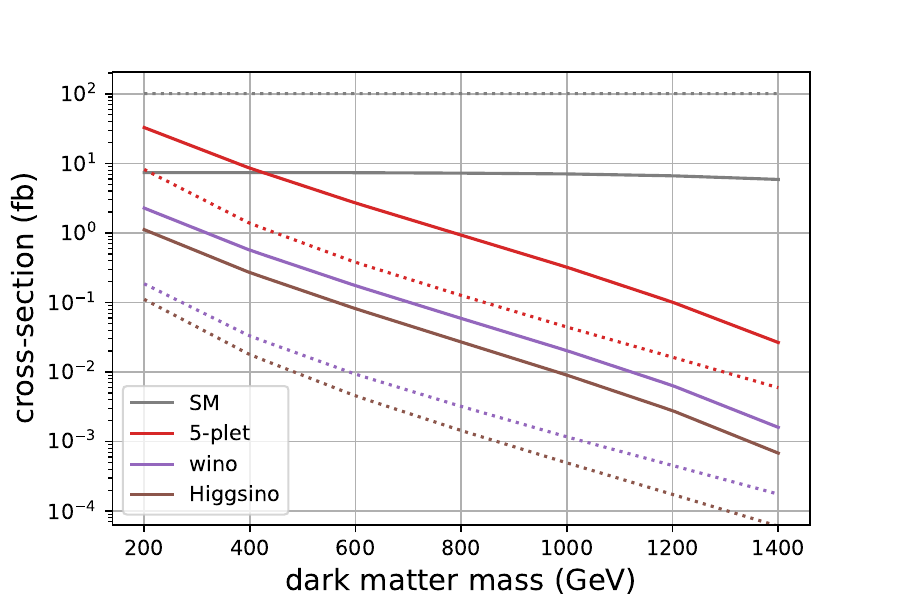}
    \caption{The cross section of the standard model backgrounds and the signals from the 5-plet, wino and Higgsino for $\sqrt{s} = 5\,\text{TeV}$ after cuts discussed in the main text. The solid lines are for the mono-$\mu$ channel and the dotted lines are for the mono-$\gamma$ channel. The gray, red, purple, and brown lines are for the standard model background, the 5-plet, wino, and Higgsino, respectively.}
    \label{fig:direct_bgfg}
\end{center}
\end{figure}

In Fig.\,\ref{fig:direct_bgfg}, we show the cross sections after the cuts for the standard model backgrounds and the signals from the Majorana 5-plets, wino and Higgsino for the mono-$\mu$ and mono-$\gamma$ channels for $\sqrt{s} = 5\,\text{TeV}$. The WIMPs are produced via the vector boson fusion (VBF) process, as shown in Fig\,\ref{fig:signal_monomu} and Fig.\,\ref{fig:signal_monophoton}, and the cross sections scale as approximately $sm^{-4}$\,\cite{Han:2020uak}.
We find that the signal cross sections of the mono-$\gamma$ channel are much smaller than the corresponding mono-$\mu$ channel cross sections whereas the mono-$\gamma$ background cross section is larger. 
For higher multiples, such as the 5-plets, the production cross sections are generically large and we expect better sensitivities for larger mass regions. At $\mu^+\mu^-$ colliders, the DY processes have a large cross section for such larger mass regions. This is the reason why Ref.\,\cite{Han:2020uak} concludes that the mono-$\gamma$ channel is effective for the higher multiples. However, the lack of the DY process at $\mu^+\mu^+$ collider spoils the advantage of the mono-$\gamma$ channel over the mono-$\mu$ channel; even though the background cross sections for the mono-$\gamma$ channel at $\mu^+\mu^+$ colliders are much suppressed than that at $\mu^+\mu^-$ colliders, still it is larger than that of the mono-$\mu$ channel after the cuts and the signal cross sections are smaller. Thus, we conclude that the mono-$\mu$ channel is the best discovery mode of the WIMP direct production at $\mu^+\mu^+$ colliders. For the range of $\sqrt{s}$ which we investigate, the background cross section of the mono-$\mu$ channel scales as $\sim 1 / s$ and is $\mathcal{O}(1\mbox-10)\,\text{fb}$. Thus, the $\mathcal{O}(0.1)\,\%$ systematic uncertainty does not affect much if the integrated luminosity is $\lesssim\mathcal{O}(10)\,\text{ab}^{-1}$, as reported in Ref.\,\cite{Han:2020uak}.

\section{Analysis}
\label{sec:result}

\subsection{Statistical method}
In this section, we discuss the statistical method for the analysis of both the direct and indirect search. For the direct production search, we consider the mono-$\mu$ channel and the significance is estimated as $S/\sqrt{B}$, where $S$ and $B$ are the numbers of signals and backgrounds, respectively. We do not take into account the systematic uncertainties in the analysis of the direct search because the mono-$\mu$ channel is robust against systematic uncertainties as discussed in Sec.\,\ref{sec:direct}.

For the indirect search, the shape analysis is adopted. As is discussed in Sec.~\ref{sec:indirect}, the WIMP affects the angular distribution of $\mu^+\mu^+$ elastic scattering through the gauge boson propagator, and we can search for the WIMP by analyzing the angular distribution of the differential cross section. For the statistical test, we perform the binned likelihood method on the differential cross section with respect to the scattering angle $\theta$. We use 15 uniform intervals of the angle $\theta$, which satisfies $0\leq\eta\leq2.5$. Then the $\chi^2$ is obtained as
\begin{equation}
    \chi^2 = \sum_{i\in \textrm{bin}}\frac{\qty(N^{(SM+WIMP)}_i - N^{(SM)}_i)^2}{N^{(SM)}_i+(N^{sys}_i)^2},
    \label{eq:chi}
\end{equation}
where $N^{(SM)}_i\ \qty(N^{(SM+WIMP)}_i)$ is the SM (SM+WIMP) prediction of the number of events in the bin $i$. In the above expression, $N^{sys}_i$ is the systematic error in the bin $i$, and $N^{(SM)}_i$ in the denominator is the statistical error squared.

Although the signals at lepton colliders are clean, many sources cause systematic errors at the collider experiment, like luminosity, angular resolution, etc~\cite{Harigaya:2015yaa}. The precise estimation of the systematic error needs details of the detector performance and the experimental setup, which is beyond our scope. In this article, the systematic error is assumed to be universal over the bins; the systematic error of the number of events in the bin $i$ is given by $N^{sys}_i = \epsilon N_i^{SM}$. We consider the $\mathcal{O}(0.1\%)$ systematic error for each bin, $\epsilon\in[0\%,0.3\%]$. The value, $\mathcal{O}(0.1)$\,\%, is an expected order of the systematic error at the future lepton collider. For example, Let us consider the angular resolution. FCC-ee proposal estimates the angular resolution for high energy muon around $0.3$ mrad in the barrel region \cite{Bacchetta:2019fmz}. When we use the 15 intervals in the multi-bin analysis, this angular resolution causes less than $0.1$ \% systematic error at each bin. 

\begin{figure}[t]
\begin{center}
    \centering
    \begin{tabular}{cc}
    
    \begin{minipage}{0.5\hsize}
        \centering
        \includegraphics[scale=0.6]{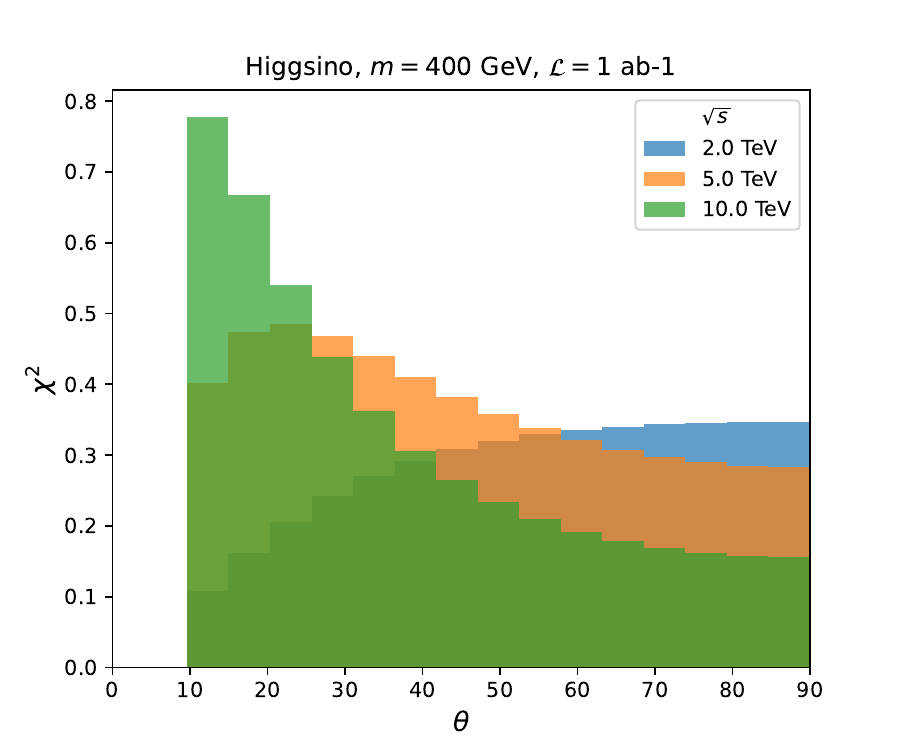}
    \end{minipage}
    
    \begin{minipage}{0.5\hsize}
        \centering
        \includegraphics[scale=0.6]{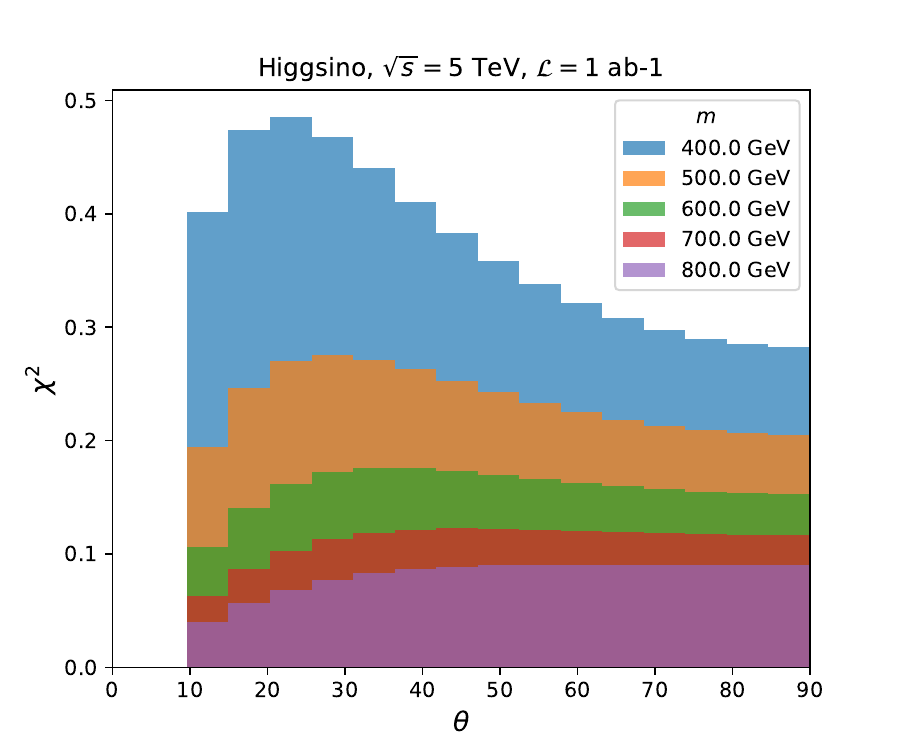}
    \end{minipage}
    
    \end{tabular}
    \caption{The contribution to $\chi^2$ from each bin with fixed WIMP mass (left), and fixed $\sqrt{s}$ (right). The Higgsino search with luminosity $\mathcal{L}=1\ \textrm{ab}^{-1}$, the muon polarization $P_{\mu}=0$, and the systematic error $\epsilon=0.3$ \% is considered. The angular cutoff is $\eta=2.5$.}
    \label{fig:chi_angular}
\end{center}
\end{figure}

Let us discuss the behavior of $\chi^2$ when we vary $\sqrt{s},\ m$, and $\mathcal{L}$.
We first examine the contribution to $\chi^2$ from each bin.
Approximating that the interval of the bins, $\dd\theta$, is much smaller than unity, the contribution to $\chi^2$ from a bin spanning $\theta$ to $\theta + \dd \theta$, $\dd \chi^2$, can be written as
\begin{equation}
    \frac{\dd \chi^2}{\dd \theta} = \left. \qty(\mathcal{L}\frac{\dd \sigma_\text{WIMP}}{\dd \theta})^2\middle / \qty(\mathcal{L}\frac{\dd \sigma_\text{tree}}{\dd \theta})\right. = \mathcal{L}\Delta^2(\theta)\frac{\dd \sigma_\text{tree}}{\dd \theta},
    \label{eq:chi_ang}
\end{equation}
where $\mathcal{L}$ is the integrated luminosity. For simplicity, we ignore the systematic errors here.

Let us consider the case of $s \gg m^2$; in such a case, the forward scattering contributes to $\chi^2$ the most.
In the forward region ($\cos\theta\sim 1$), the $t$-channel process dominates $\dd \sigma_{\text{WIMP}}/\dd \theta$.  Using the fact that $\Delta(\theta)$ is proportional to $\Pi(t/m^2)$ there and that $\frac{\dd \sigma_{\text{tree}}}{\dd \cos\theta}\propto s/t^2$, Eq.~\eqref{eq:chi_ang} approximately gives
\begin{align}
    \frac{\dd \chi^2}{\dd \theta} \propto \frac{\mathcal{L}\sqrt{s}\cos\frac{\theta}{2}}{t^{3/2}}\qty[\int \dd y y(1 - y) \log\qty(1-\frac{t}{m^2}y(1 - y))]^2.
    \label{eq:chi_apr}
\end{align}
Eq.~\eqref{eq:chi_apr} indicates that $d\chi^2/d\theta$ has a peak at $\theta\sim 2\sin^{-1} (m/\sqrt{s})$, which corresponds to $t\sim -m^2$. 
The height of the peak is expected to be approximately proportional to $\sqrt{s}/m^3$. The discussion above assumes only the statistical error, but small enough systematic errors are expected not to change the qualitative behavior of $\chi^2$.
In Fig.~\ref{fig:chi_angular}, we show the contributions of each bin to $\chi^2$ in our analysis for the Higgsino search with a fixed value of the WIMP mass (left) and the center-of-mass energy $\sqrt{s}$ (right). As expected, there is a peak around $t\sim -m^2$, and the height of the peak is roughly proportional to $\sqrt{s}/m^3$, although $u$-channel contribution blurs the relation. 

\begin{figure}[t]
\begin{center}
    \centering
    \begin{tabular}{cc}
    
    \begin{minipage}{0.5\hsize}
        \centering
        \includegraphics[scale=0.5]{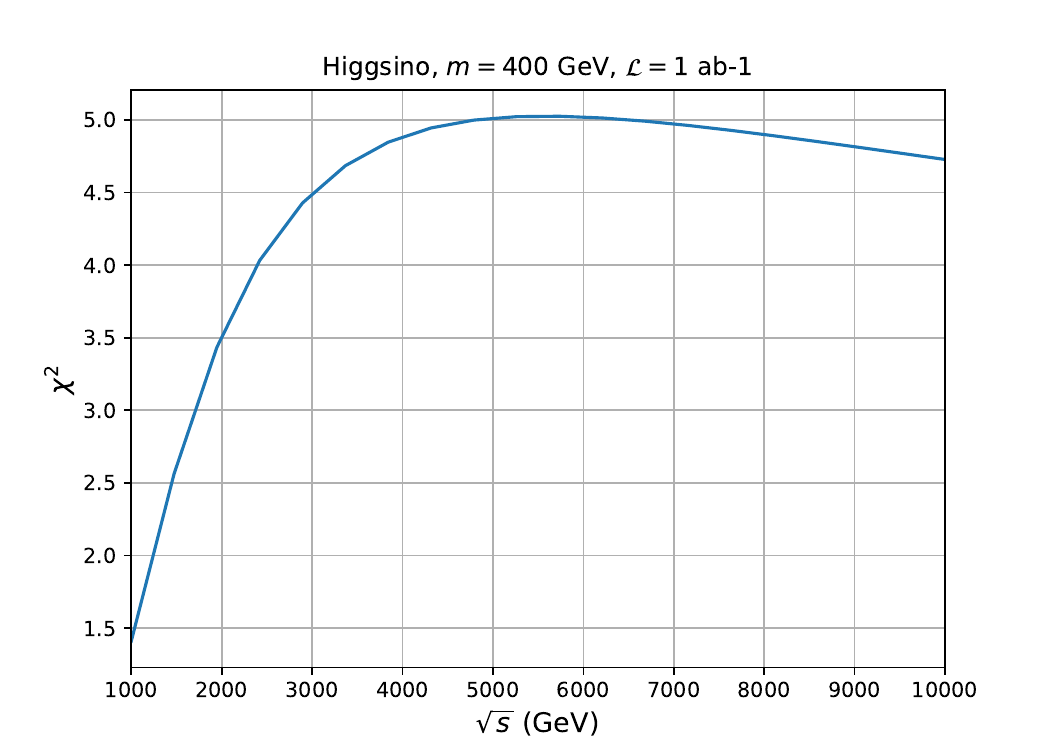}
    \end{minipage}
    
    \begin{minipage}{0.5\hsize}
        \centering
        \includegraphics[scale=0.5]{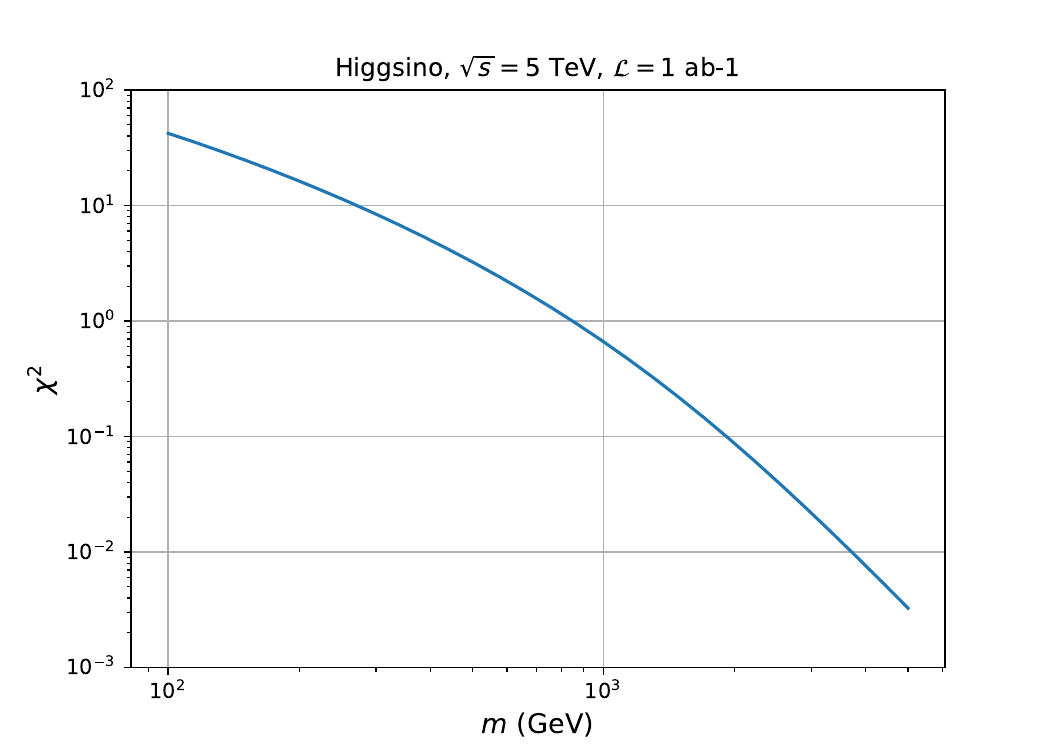}
    \end{minipage}
    
    \end{tabular}
    \caption{The center of mass energy dependence (left) and the WIMP mass dependence (right) of the chi-squared value. The setting is the same as Fig.~\ref{fig:chi_angular}.}
    \label{fig:chi_dependence}
\end{center}
\end{figure}

Let us go back to $\chi^2$. $\chi^2$ is the sum of the contributions from each bin; approximately, it is obtained by 
integrating $\dd \chi^2/\dd \theta$ over 
$\theta \in [\delta, \pi / 2]$, where $\theta = \delta$ corresponds to $\eta = 2.5$. With the WIMP mass being fixed, the width of the peak of $d\chi^2/d\theta$ is proportional to $1/\sqrt{s}$ whereas the height is proportional to $\sqrt{s}$.
Therefore, in the limit of large $s$, $\chi^2$ becomes insensitive to $s$ as far as the peak position is within $[\delta, \pi / 2]$. With a non-zero $\delta$, the position of the peak of $\dd \chi^2/\dd \theta$ eventually becomes smaller than $\delta$ for large $s$ and $\chi^2$ is reduced.
On the other hand, for $s \sim \mathcal{O}(10 m^2)$, the width of the peak width is broad as we see in Fig.\,\ref{fig:chi_angular}. The magnitude of $\chi^2$ is determined by the height of the peak and roughly proportional to $\mathcal{L} \sqrt{s}/m^3$. In Fig.~\ref{fig:chi_dependence}, we show $\chi^2$ of our analysis for Higgsino. In the left figure, we fix $m = 400\,\text{GeV}$ and vary $\sqrt{s}$. We indeed see $\chi^2$ approaches to the constant value and reduces for large $\sqrt{s}$. In the right figure, we fix $\sqrt{s} = 5\,\text{TeV}$ and show the mass dependence of $\chi^2$. $\chi^2$ is roughly proportional to $m^3$ for $s / m^2 \sim 10$ as expected.

\subsection{Result}
We now evaluate the 95\,\% C.L. discovery reach for the WIMP at $\mu^+\mu^+$ collider as a function of $\sqrt{s}$.
We consider the Higgsino ($n=2,\ Y=\frac{1}{2}$, Dirac fermion), the wino ($n=3,\ Y=0$, Majorana fermion) and the 5-plet minimal dark matter ($n=5,\ Y=0$, Majorana fermion). The initial $\mu^+$ is assumed to be either the polarized beam ($P_{\mu^+}=0.8$) or the unpolarized beam ($P_{\mu^+}=0$).

\begin{figure}[t]
\begin{center}
    \centering
    \begin{tabular}{cc}
    
    \begin{minipage}{0.5\hsize}
        \centering
        \includegraphics[scale=0.5]{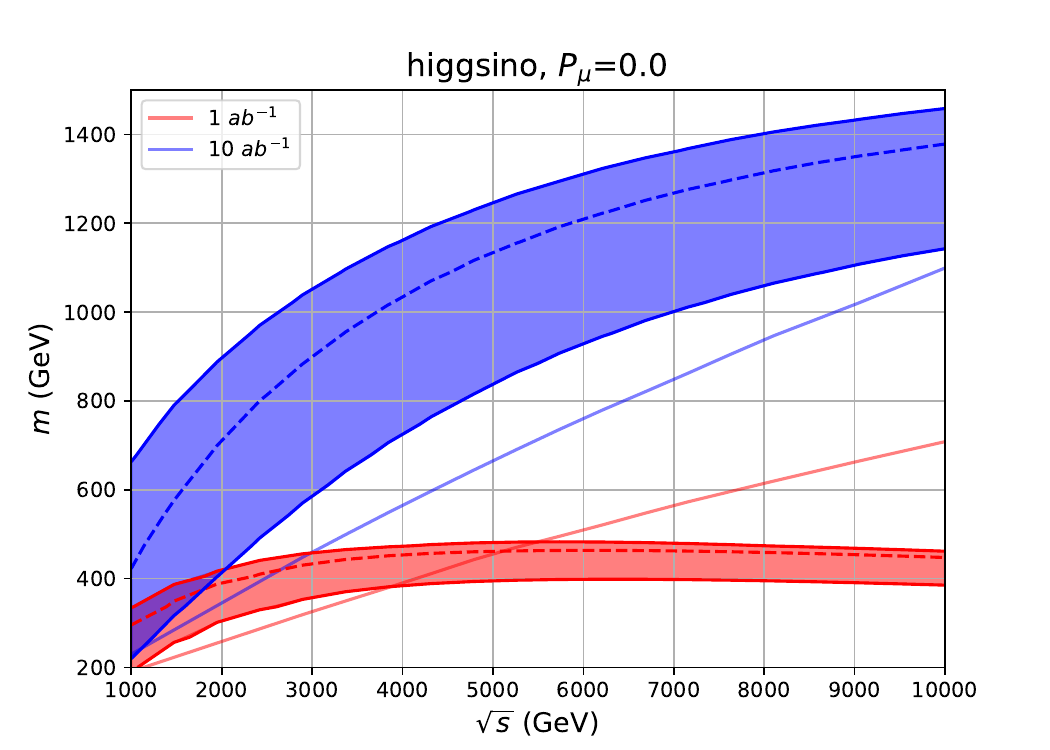}
    \end{minipage}
    
    \begin{minipage}{0.5\hsize}
        \centering
        \includegraphics[scale=0.5]{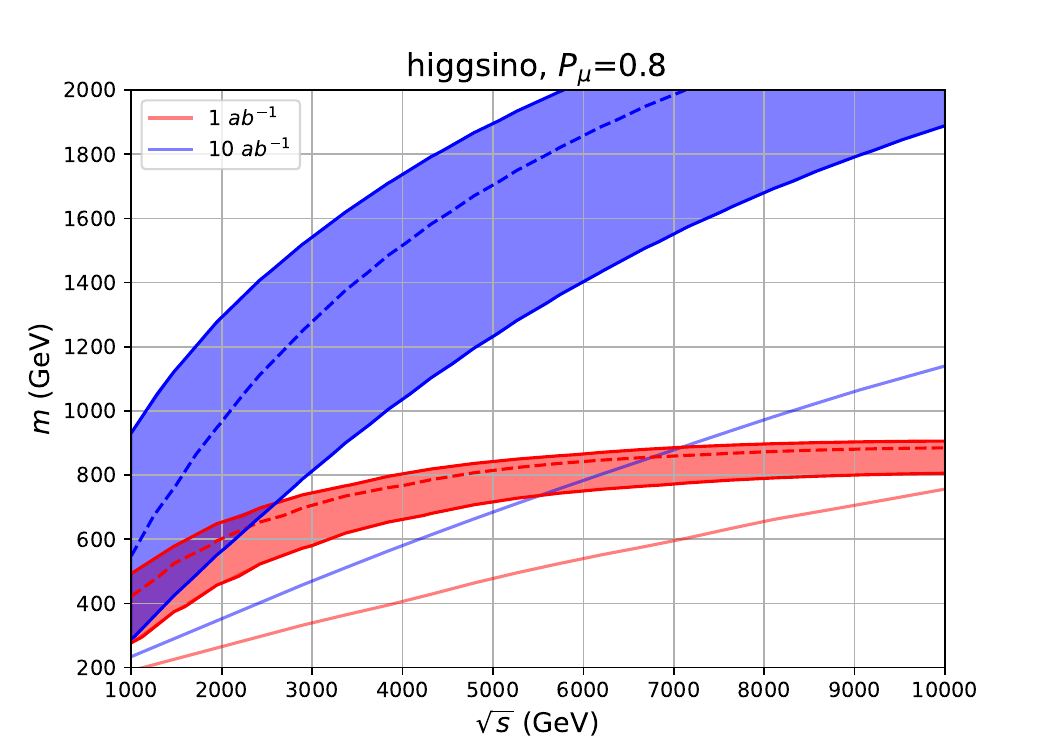}
    \end{minipage}
    
    \end{tabular}
    \caption{The discovery reach for the Higgsino by the indirect search (band) and the direct production search (solid line). The systematic error is assumed to be $\epsilon\in[0\%,0.3\%]$, and the discovery reach with $\epsilon=0.1$ \% is described by the dashed line. We assume that the integrated luminosity is $1\ \textrm{ab}^{-1}$ (red) and $10\ \textrm{ab}^{-1}$ (blue), and the initial muon polarization is $P_{\mu}=0$ (left panel) and $P_{\mu}=0.8$ (right panel).}
    \label{fig:higgsino}
\end{center}
\end{figure}

\begin{figure}[t]
\begin{center}
    \centering
    \begin{tabular}{cc}
    
    \begin{minipage}{0.5\hsize}
        \centering
        \includegraphics[scale=0.5]{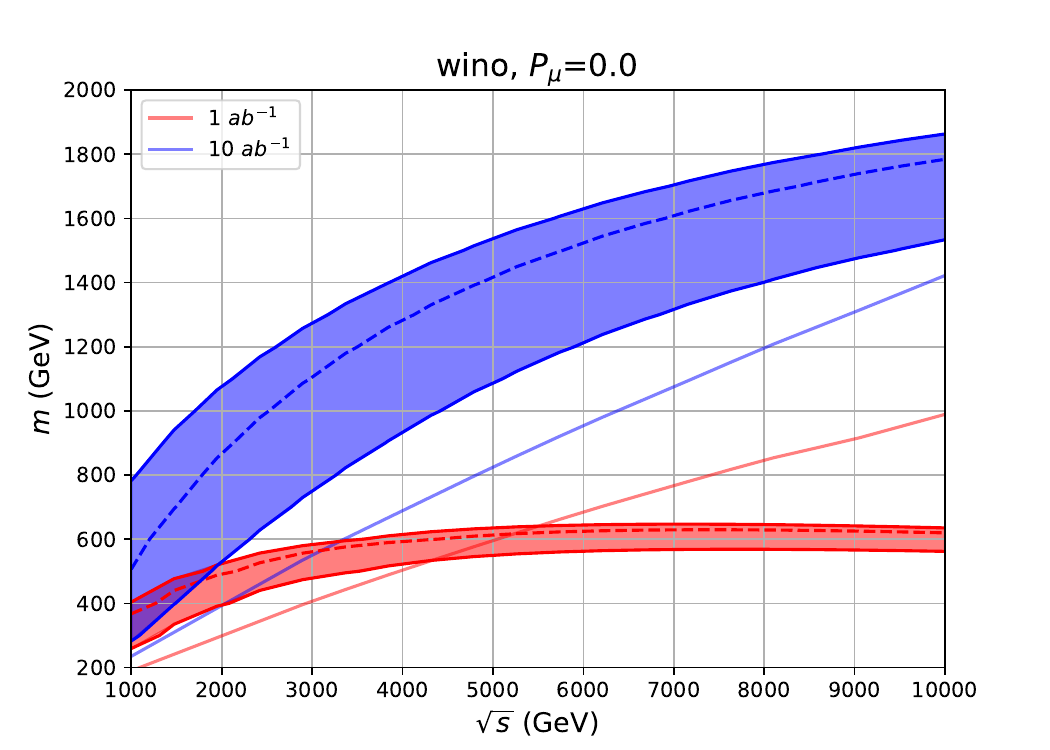}
    \end{minipage}
    
    \begin{minipage}{0.5\hsize}
        \centering
        \includegraphics[scale=0.5]{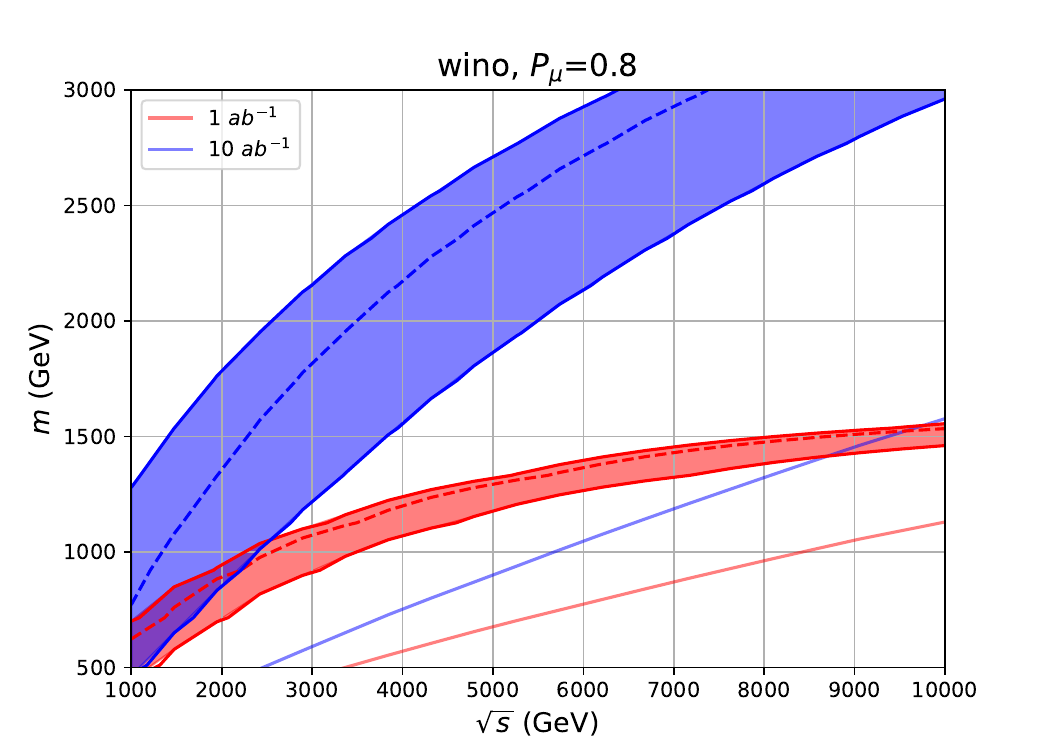}
    \end{minipage}
    
    \end{tabular}
    \caption{The discovery reach for the wino.}
    \label{fig:wino}
\end{center}
\end{figure}

\begin{figure}[t]
\begin{center}
    \centering
    \begin{tabular}{cc}
    
    \begin{minipage}{0.5\hsize}
        \centering
        \includegraphics[scale=0.5]{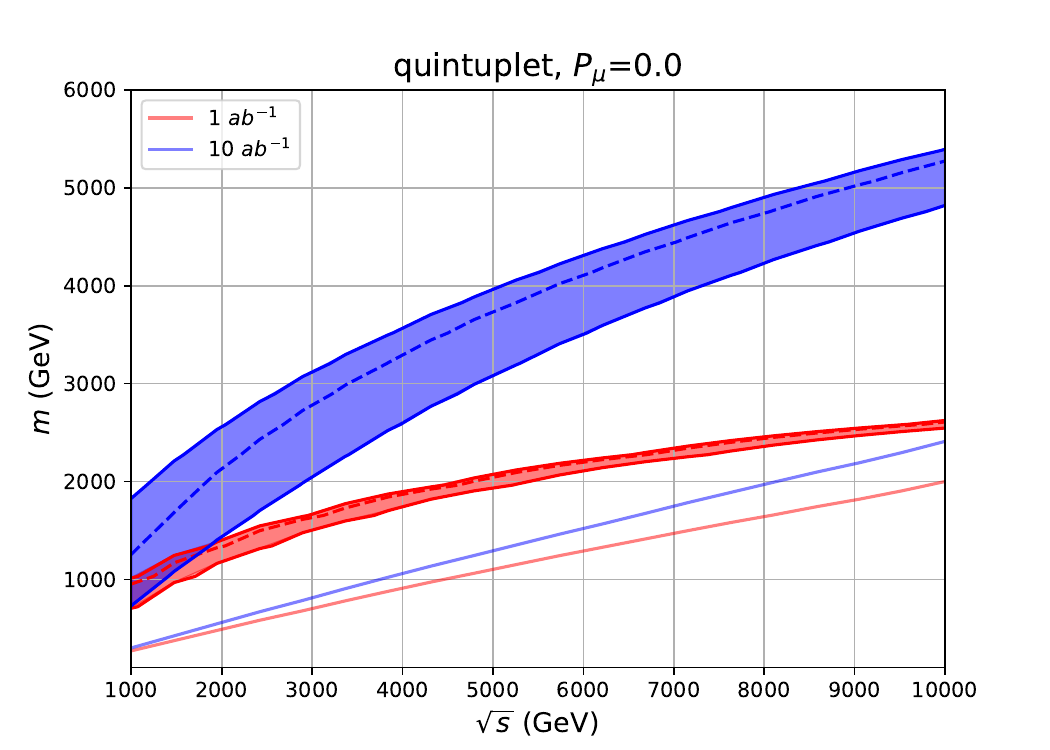}
    \end{minipage}
    
    \begin{minipage}{0.5\hsize}
        \centering
        \includegraphics[scale=0.5]{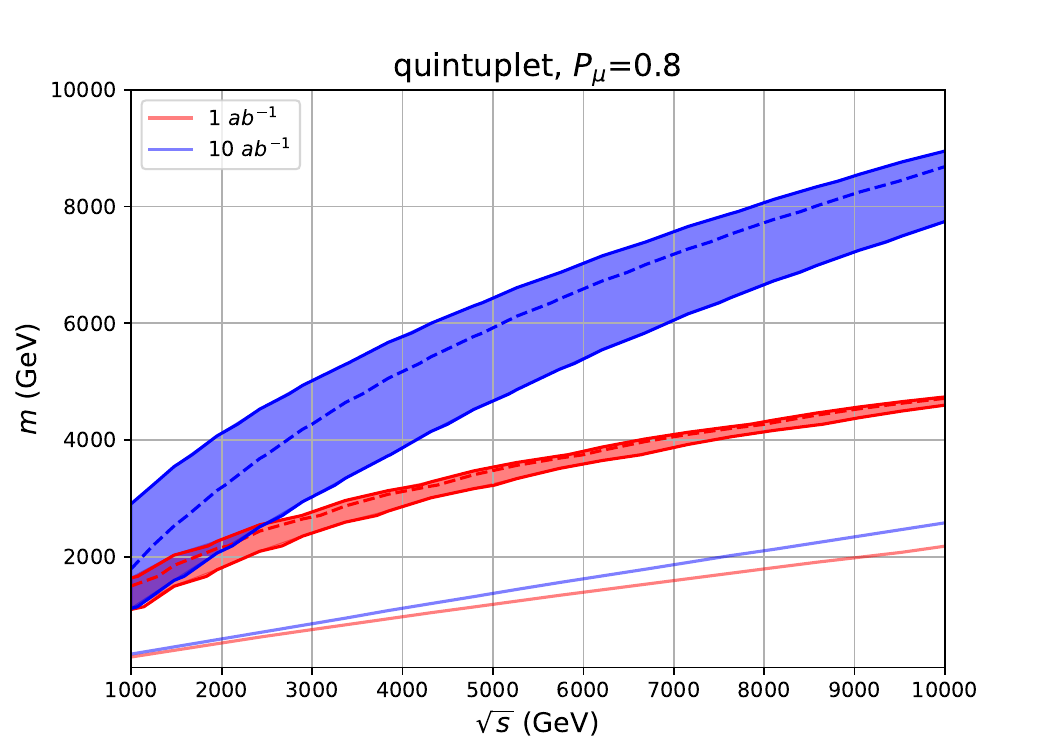}
    \end{minipage}
    
    \end{tabular}
    \caption{The discovery reach for the 5-plet fermionic minimal dark matter.}
    \label{fig:5plet}
\end{center}
\end{figure}

Figs.~\ref{fig:higgsino}, \ref{fig:wino} and \ref{fig:5plet} show the discovery reach of the Higgsino, the wino and the $5$-plet, respectively.
The bands represent the reach of the WIMP mass by the indirect search at each $\sqrt{s}$ assuming the universal systematic error $\epsilon\in[0\,\%,0.3\,\%]$. The upper (lower) lines correspond to $0\,\%$ ($0.3\,\%$) systematics. We show the bound with $0.1\,\%$ systematics in the dashed lines. For comparison, we also show the discovery reach of the mono-$\mu$ direct production channel as the solid lines. The integrated luminosity $\mathcal{L}$ is assumed to be $1\ \textrm{ab}^{-1}$ (red), and $10\ \textrm{ab}^{-1}$ (blue). The left panels are for the unpolarized initial muons, $P_{\mu^+}=0$, and the right panels are for the polarized initial muons, $P_{\mu^+}=0.8$.

Let us compare the indirect detection with the direct production channel. For a small luminosity, generally speaking, the expected sensitivities of the indirect detection are better than the direct production for small $\sqrt{s}$, whereas the direct production has an advantage for large $\sqrt{s}$. As discussed in the previous subsection, $\chi^2$ for the indirect detection does not grow for sufficiently large $\sqrt{s}$. On the other hand, for $m^2 \ll s$, the direct production cross section scales as $s/m^4$ and the background cross section scales as $1 / s$. In total, the significance of the direct production scales as $s^{3/2} / m^4$. Therefore, the direct production process has an advantage for large $s$ and small $m$. 

As the luminosity increases, the advantage of the indirect detection process over the direct production process becomes clearer. This is because the significance of the indirect production scales as $\mathcal{L}m^3$ for $s \sim \mathcal{O}(10 m^2)$ as we have discussed in the previous section, whereas the significance of the direct production scales as $\sqrt{\mathcal{L}}/m^4$. 
We thus conclude that the indirect detection is more capable for a large integrated luminosity, $\sim 10\,\text{ab}^{-1}$.
With a positive beam polarization $P_{\mu^+} > 0$, both the signal and background cross section increase, although the signal cross section increases more. Therefore, the initial beam polarization is effective for the indirect search, as we can see in the right panels of the figures.

In Fig.~\ref{fig:higgsino}, we can see that the Higgsino mass suggested by the thermal WIMP scenario is covered for $10\ \text{ab}^{-1}$ integrated luminosity. As the thermal mass target of the Higgsino is relatively low, it can be probed by both the indirect and direct search, but the indirect search can cover it with lower $\sqrt{s}$. With the polarized beams, the thermal mass target is covered with $\sqrt{s} = 2\,\text{TeV}$ and $0.1\,\%$ systematic uncertainties. For the wino, as the thermal mass target is heavier, only the indirect search can probe it with $\sqrt{s}$ and the integrated luminosity which we consider. From Fig.~\ref{fig:wino}, we can see that the thermal mass target is covered with $\sqrt{s} = 6\,\text{TeV}$ and $0.1\,\%$ systematic uncertainties. For the wino and $5$-plet, the mass coverage of the indirect process may be larger than $\sqrt{s} / 2$ for $10\ \text{ab}^{-1}$ integrated luminosity with good accuracy.

\section{Conclusion and discussion}
\label{sec:summary}
In this study, we have investigated the prospect of probing the WIMP at the $\mu^+\mu^+$ colliders. Since the WIMP affects the vacuum polarization of the electroweak gauge bosons, the angular distribution of the $\mu^+\mu^+$ M{\o}ller scattering is altered by the existence of the WIMP. We have focused on the precise measurement of the $\mu^+\mu^+$ M{\o}ller scattering and estimated the capability of the $\mu^+\mu^+$ collider to search for the WIMP by this indirect search. For the comparison with the indirect search, we have also discussed the capability of the $\mu^+\mu^+$ collider by the direct production of the WIMP. We have considered the mono-$\mu$ channel and the mono-$\gamma$ channel and found that the mono-$\mu$ channel gives the most severe constraint on the WIMP at the $\mu^+\mu^+$ collider.

Our main results are shown in Figs.~\ref{fig:higgsino},~\ref{fig:wino}, and~\ref{fig:5plet}. We show that the indirect search can probe the thermal mass target of the Higgsino and the wino at $\mathcal{O}(1)$ TeV $\mu^+\mu^+$ collider with $1-10\ \textrm{ab}^{-1}$ luminosity.
In addition, the indirect search outperforms the direct production search with a large luminosity, that is, the indirect search achieves a broader discovery reach for the WIMP across a wide range of $\sqrt{s}$, various types of the WIMPs, and different beam polarizations. This is due to the weaker WIMP mass dependence of the indirect search sensitivity. The advantage of the indirect search holds even if the systematic uncertainty extends to 0.3\%. Importantly, these results hold true for any new particle with $\text{SU}(2)_L\times\text{U}(1)_Y$ charges, extending the applicability of our findings.

We assume that the systematic error is universal over the bins and that it is of $\mathcal{O}(0.1)$ \%. As shown in Sec.~\ref{sec:result}, the discovery reach by the indirect search is sensitive to the systematic error and the precise estimation of the systematic errors in the muon M{\o}ller scattering is essential to consider the capability of the $\mu^+\mu^+$ collider. Careful consideration of experimental errors, such as luminosity measurements and detector resolutions, alongside theoretical uncertainties like photon emissions, is vital in accurately assessing the $\mu^+\mu^+$ collider's capabilities.

While we have stressed the capability of the indirect detection, the direct production approach remains robust against systematic uncertainties and serves as an independent mode of investigation. We acknowledge that the applied cuts, inherited from previous $\mu^+\mu^-$ collider studies\,\cite{Han:2020uak}, can be optimized, especially for the mono-$\gamma$ channel in $\mu^+\mu^+$ colliders. Furthermore, although challenging due to significant beam-induced backgrounds, the possibility of a disappearing track search in $\mu^+\mu^+$ colliders, akin to $\mu^+\mu^-$ colliders\,\cite{Capdevilla:2021fmj}, offers a promising avenue for future exploration.
We leave these topics for future studies\,\cite{muonproj?}. 

Lastly, we extend our discussion to $\mu^+e^-$ colliders\,\cite{Hamada:2022mua}. Our study on the radiative correction due to the WIMPs can be applied to the $\mu^+e^-$ elastic scattering almost in parallel. $\mu^+e^-$ colliders generically have a greater integrated luminosity but the center-of-mass energy is smaller. As we have discussed, a small $\sqrt{s}$ can be compensated by a large integrated luminosity for the indirect search, and the indirect search may have an advantage over direct production searches. Because 
the muon energy is assumed to be much larger than the electron energy and the collision is asymmetric for $\mu^+e^-$ colliders, the sensitivities for the forward scattering may be spoiled, but it can be an interesting alternative.

\acknowledgments
TM was supported by JSPS KAKENHI Grant (No.\ 22H01215).
AN was supported by JSPS KAKENHI Grant (No.\ 22J21016). SFW was supported by the Global Science Graduate Course program of the University of Tokyo and receive financial support from Daikin Industries, Ltd.



\bibliography{papers}

\end{document}